\documentclass[prper,longbibliography,reprint,floatfix,superscriptaddress]{revtex4-2}
\usepackage[T1]{fontenc}	% should generally be included for better accented-word behavior
\usepackage[latin9]{inputenc}	% should generally be included for better accent behavior
\usepackage{geometry}		% for controlling page margins
\usepackage{graphicx}
\usepackage[above,below]{placeins}	% allows use of \FloatBarrier command to force section barriers
\usepackage{times}
\usepackage{xcolor}
\usepackage{textgreek}
\usepackage{amsmath,amssymb}
\usepackage[italic]{mathastext}
\usepackage{outlines}
\usepackage{subfiles}
\usepackage{caption}
\usepackage{subcaption}

\usepackage{hyperref}  
\hypersetup{colorlinks=true,urlcolor=blue,citecolor=blue,linkcolor=blue}   
\usepackage{enumitem}          % package for handling list formatting
\setlist{nosep}                 % Tightest spacing for lists. `noitemsep` is more relaxed

\begin{document}

%\begin{titlepage}

\title{A framework for characterizing covariational reasoning in physics}

\author{Alexis Olsho}
\affiliation{Department of Physics and Meteorology, United States Air Force Academy, 2354 Fairchild Drive, USAF Academy, CO 80840 USA}

\author{Charlotte Zimmerman}
\affiliation{Department of Physics, University of Washington, Box 351560, Seattle, WA 98195-1560, USA}

\author{Suzanne White Brahmia}
\affiliation{Department of Physics, University of Washington, Box 351560, Seattle, WA 98195-1560, USA}

\begin{abstract}
Covariational reasoning---considering how changes in one quantity affect another, related quantity---is a foundation of quantitative modeling in physics. Understanding quantitative models is a learning objective of introductory physics instruction at the college level. Prior work suggests that covariational reasoning in physics contexts differs in important ways from reasoning about functions and graphs in purely mathematical contexts; this reasoning is effortful in physics even for mathematically well-prepared students. In order to help students learn to reason covariationally in physics contexts, we need to characterize what we mean by physics covariational reasoning. To this end, we present a framework of covariational reasoning in physics contexts, to describe the ways that covariational reasoning is used in physics modeling. The framework can be used as a tool by which instructors can recognize physics covariational reasoning patterns and researchers can analyze student reasoning. The framework can also help inform the development of effective instructional materials and methods.

\end{abstract}

\maketitle

\section{Introduction}

In a typical physics lecture, statements such as ``\dots it goes like $1/r$\dots'' or ``\dots they're inversely proportional\dots'' are common, and often considered part of  ``thinking like a physicist.'' More specifically, these statements are examples of \emph{covariational reasoning}---considering how changes in one quantity affect another, related quantity \cite{carlson2002,thompson2017}. Covariation of quantities is central to reasoning in physics, especially reasoning related to quantitative modeling. Given how productive this reasoning is in physics, we would like our students to leave our courses with a facility for covariational reasoning. However, unlike well established content-based learning objectives (such as Newton's Laws or conservation of energy), effective methods for helping students learn to use covariational have not been identified. In addition,
instructors may not recognize a need for instruction in covariational reasoning, and indeed often expect their students to come into class able to reason this way from experiences in prerequisite mathematics courses.

While many students do enter physics courses with some covariational reasoning experience from previous courses in mathematics, research has demonstrated that some important types of reasoning learned in mathematics courses do not translate directly to physics contexts \cite{redish2015,kuo2013,Hu2013}. In addition, introductory physics courses may not be succeeding at helping students learn this kind of reasoning. Recent research into measuring physics students' covariational reasoning finds that it does not improve substantially as a result of instruction in introductory-level physics, even in settings where the instruction is research-based \cite{white2021PIQL}. Understanding more fully how covariational reasoning is used in physics will benefit both researchers and instructors. 

Covariational reasoning frameworks developed by mathematics education researchers have served as a guide for the creation of precalculus curricular materials that show promise towards helping students learn this kind of thinking \cite{carlson2002, pathwaysResearch}.
These frameworks have also been a productive lens in mathematics education research to identify important ways that students and experts reason. While covariation is a relatively new research lens in physics education, early results suggest the covariational reasoning frameworks from mathematics education research are productive for analyzing physics students' work, but do not fully characterize the ways covariation is used in physics \cite{Emigh2019, VanDenEynde2019, expertcovar}. 

Based on these findings, we suggest that \emph{physics covariational reasoning} is distinct from covariational reasoning as taught in mathematics courses, and that physics courses aren't currently very successful at helping students learn to reason in this way. As a first step in helping improve learning outcomes, we propose that physics covariational reasoning be characterized in an assessable, fine-grained way to inform subsequent instructional materials and methods designed to help physics students develop covariational reasoning.

Using prior work by both mathematics and physics education researchers, we have developed the Covariational Reasoning in Physics (CoRP) framework to formally characterize physics covariational reasoning. The CoRP framework operationalizes physics covariational reasoning, distinguishing it from covariational reasoning as described in the mathematics education literature. The CoRP framework is informed by observations of physics experts engaged in covariational reasoning, as well as prior work in both physics and mathematics education research. We designed the CoRP framework as a tool for physics instructors and physics education researchers to catalyze change, organizing a complex phenomenon into one that can be comprehended more readily and usefully.  

In this paper, we describe the development and relevance of the CoRP framework. We demonstrate its use for instructors and researchers as a tool to analyze students' covariational reasoning, which can identify topics for instruction and aid in the develop targeted interventions.

\section{Background}
\label{sec:CRbackground}

Our characterization of physics covariational reasoning is built on work by mathematics education researchers, and is informed by both established and more recent work in physics education research. In this section, we describe the frameworks of covariational reasoning developed by mathematics education researchers. We then discuss research findings from physics education that provide some structure for experts' covariational reasoning in physics. Finally, we describe the building blocks of mathematical modeling that are essential to covariational reasoning and that appear directly in the CoRP framework.

\subsection{Covariational reasoning in mathematics education research}

\emph{Covariational reasoning} has been defined by mathematics education researchers as ``the cognitive activities involved in coordinating two varying quantities while attending to the ways in which they change in relation to each other'' \cite{carlson2002}; that is, it describes reasoning about how changes in one quantity affect changes in another, related quantity. In mathematics education research, covariational reasoning has been studied widely and has been identified as an essential part of reasoning in pre-calculus and calculus \cite{Thompson1994a, Carlson2010calc, Jones2013, BYERLEY2019, Ely2019, Boyce2021}. Covariation has been studied in contexts of reasoning about function \cite{confrey1995, Carlson1998, oehrtman2008, Johnson2016, thompson2017, Paoletti2018}, as well as graphing and the use of coordinate systems \cite{Moore2013a, Weber2014, Johnson2015, Hobson2017, Byerley2017}. It has also been identified as necessary for reasoning about rates of change \cite{Thompson2011a}. 

In 2002, mathematics education researchers Carlson, Jacobs, Coe, Larsen, and Hue developed frameworks describing hierarchical levels and associated ``mental actions'' (MA) of covariational reasoning \cite{carlson2002}, based on studies of undergraduate math students interpreting and creating representations of functions \cite{Carlson1998}. The covariational reasoning mental actions were designed to allow researchers and educators to assess the level of students' covariational reasoning. Each of the mental actions is associated with specific behaviors related to covariational reasoning. The mental actions range from a recognition that variables are related (MA~1), to considering the specific relationship between the variables, including the rate of change and the rate of the rate of change (MA~5). In 2017, mathematics education researchers released an updated framework of covariational reasoning \cite{thompson2017}, incorporating research performed subsequent to the development of the original framework. Table \ref{tab:mathcovar} shows a summary of the relevant aspects of the 2002 and 2017 covariational reasoning frameworks, adapted from Jones \cite{Jones2022MultivariationReasoning}. 

\renewcommand{\arraystretch}{1.5}
\setlength{\tabcolsep}{6pt}
\begin{table*}
\begin{ruledtabular}

    \centering
    \begin{tabular}{p{0.05\textwidth} p{0.22\textwidth} p{0.3\textwidth} p{0.35\textwidth}}
       % \toprule
         Label & Mental Action \cite{carlson2002,thompson2017} & Brief Description \cite{Jones2022MultivariationReasoning} & Example Behavior \\
         \toprule
         MA~1 & Recognize Dependence & Identify variables that are dependent & Labeling axes \\
         MA~1.5 & Precoordination & Asynchronous changes in variables & Articulating that first, one quantity changes, and then the other changes \\
         MA~2 & Gross Coordination & General increase/decrease relationship & Describing that as one quantity increases, another decreases \\
         MA~3 & Coordination of Values & Tracking variable's values & Plotting points \\
         MA~4 & Chunky Continuous & Values changing in discrete chunks & Articulating that as one quantity doubles, the other triples \\
         MA~5 & Smooth Continuous & Continuous, simultaneous changes & Describing that the quantities vary together, smoothly and continuously\\
        % \bottomrule
    \end{tabular}
    \caption{A summary of the covariational reasoning mental actions (MA) frameworks developed by mathematics education researchers \cite{carlson2002,thompson2017}. Summary adapted from Jones \cite{Jones2022MultivariationReasoning}.}
    \label{tab:mathcovar}
 \end{ruledtabular}
 \end{table*}

\subsection{Covariational reasoning in physics education research}

In physics education research, reasoning about how two or more quantities change with respect to one another often falls under the names of proportional reasoning or scaling \cite{HankyPanky, Boudreaux2015, boudreaux2020toward, Bissell2022}. Proportional reasoning typically refers to directly proportional relationships (i.e. $F \propto a$), and has at times been extended to refer to non-linear relationships (i.e. $U \propto -1/r$). Scaling is often used in geometric contexts; however, it is also used throughout the literature to refer to relating discrete changes of two quantities (e.g., ``if I double this quantity, what happens to that quantity?''). In the language of covariational reasoning, we consider proportional reasoning to be \emph{linear} covariational reasoning and scaling to be an instance of \emph{discrete} covariation.

\begin{figure}
     \centering
     \includegraphics[width=0.35\textwidth]{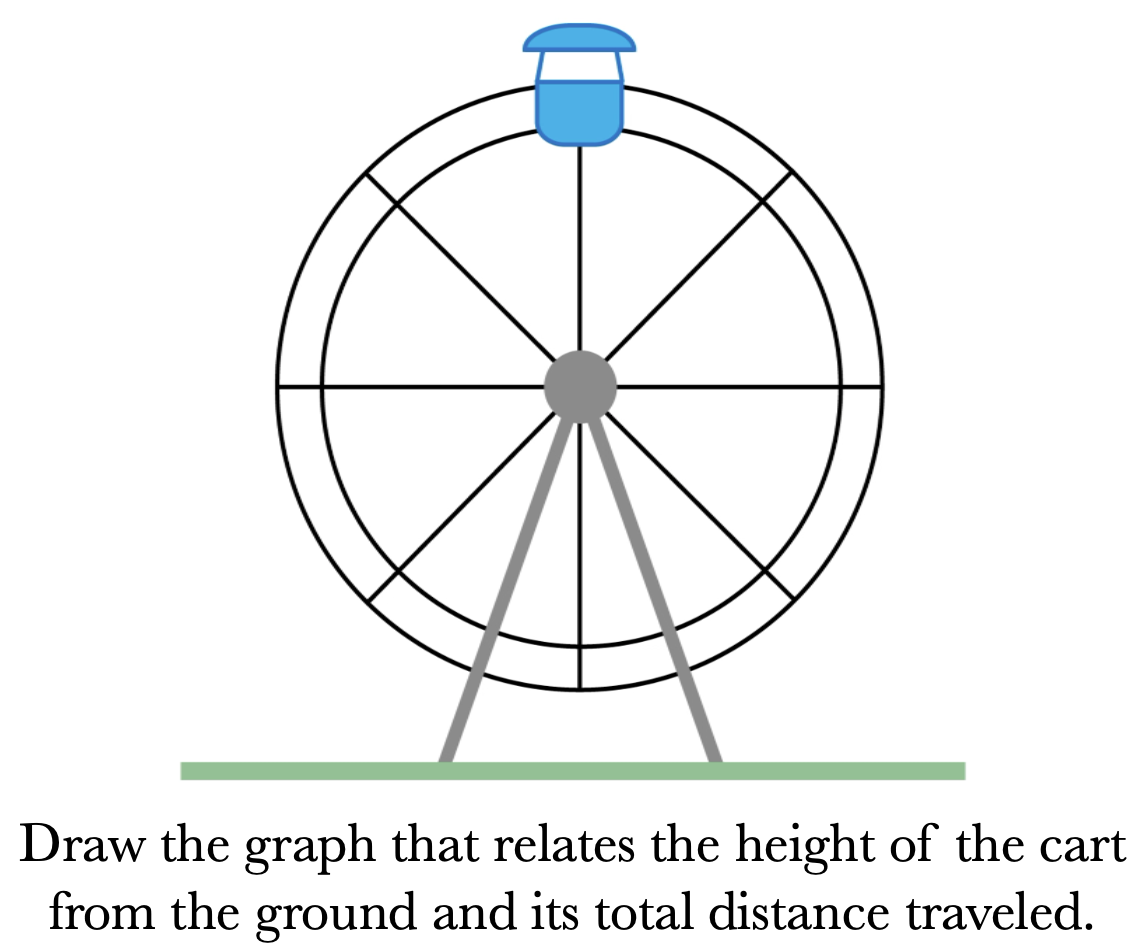}
     \caption{A still from a covariational reasoning graphing task prompt that asks experts to create a graph relating the distance traveled by a Ferris wheel cart and the height of the cart.}
     \label{fig:FerrisWheel}
 \end{figure}

Work in physics education research has demonstrated that the language of covariational reasoning from mathematics education is helpful in analyzing novice and expert work \cite{taylor2018, Emigh2019, Olsho2022, May2021, VanDenEynde2019, Zimmerman2019perc, zimmerman2020, Sokolowski2021}. Recent work by Zimmerman, Olsho, Loverude, and White Brahmia has sought to explore the extent to which the mathematics covariational reasoning mental actions framework can be used to analyze \emph{physics} covariational reasoning of physics experts engaged in modeling tasks \cite{expertcovar,strawberryfields}. The Zimmerman et al. study involved individual, think-aloud interviews with 20 physics experts (graduate students and faculty) engaged in tasks designed to elicit covariational reasoning. The tasks prompted the participants to create a graph that related two quantities. For example, one task depicted a Ferris wheel cart in motion (see Fig. \ref{fig:FerrisWheel}) and asked participants to relate the height of the cart and its total distance traveled. The overarching results of study were that:
 \begin{enumerate}
     \item Physics experts demonstrate mathematical reasoning that is consistently woven into physical sensemaking of the quantities involved \cite{strawberryfields}.
     \item Physics experts engaged in patterns of covariational reasoning and modeling that were not well described by the mathematics covariational reasoning framework summarized in Table~\ref{tab:mathcovar} \cite{expertcovar}.
 \end{enumerate}

\subsection{Foundations of physics covariational reasoning}
\label{ssec:underpinnings}

Research about the ways students and experts reason about mathematics and quantities in the context of physics underlies our work on covariation. In this section, we offer a brief overview of the fundamental ideas and terminology that may be unfamiliar to the reader but are central to our characterization of covariation in physics.

\subsubsection{Mathematization and Proceptual Understanding}
\label{sssec:PU}

\emph{Mathematization} in physics is the association of a system or context from the physical world with a mathematical representation (e.g., symbolic or graphical). One aspect of mathematization in physics is \emph{quantification}---the process of conceptualizing a system and a quality of it so that the quality includes a value, a unit of measure, and very often a sign \cite{Thompson2011a,white2019quantification}. Research in mathematics and physics education suggests that mathematization and quantification are challenging for students; for example, it has been demonstrated that students are unlikely to think of compound quantities, such as momentum, as quantities unto themselves \cite{Thompson1994a, vlassis2004, tuminaro2007egames, WhiteBrahmia2016, white2019quantification}. In addition, students likely come into physics courses with some foundational mathematical resources but may not yet use them productively for mathematization in physics \cite{Hammer2000, Boudreaux2015, white2019quantification, white2021PIQL, Odden2021}. For example, introductory physics students typically have mastered multiplying and dividing numbers; however, they may struggle to conceptualize product quantities (such as momentum and energy) and ratio quantities (such as velocity and acceleration) as distinct from the quantities that contribute to their calculation \cite{Trowbridge1980, Trowbridge1981, Boudreaux2015}.

Mathematization, including quantification, depends on connecting mathematical ideas with what they represent in the physical world. Students engaged in modeling tasks have been shown to refer back to the context of the task consistently throughout a productive modeling process as part of making sense of and validating their model \cite{Czocher2016IntroducingThinking}. \emph{Proceptual understanding} is defined by Gray and Tall as a combination of \textit{pro}cedural mastery and con\textit{ceptual} understanding \cite{gray1994}. For example, in the context of fractions, ``the symbol $\frac{3}{4}$ stands for both the process of division and the concept of fraction''; that is, a student with a proceptual understanding of fractions would move fluidly between the procedure of dividing 3 by 4, and the instantiation of the fraction $\frac{3}{4}$ as a precise quantification of portion. Maintaining a proceptual view of a mathematical representation---consistently making sense of both the mathematical formalism and the physical meaning of the expression---is a key part of modeling, and therefore essential for covariational reasoning.

Mathematics and physics meaning are interwoven in physics. This blended nature of physics and mathematics reasoning is an important facet of the body of work that describes ways in which reasoning mathematically in physics contexts is not the same as reasoning mathematically in purely mathematical contexts \cite{sherin2001, kuo2013, Hu2013, Karam2014, redish2015, WhiteBrahmia2016, Planinic2019, VanDenEynde2019}. Several researchers consider this difference through the lens of semiotics, highlighting the importance of symbols and the meaning they carry in physics \cite{redish2010makingphysics,airey2017social}. 

\subsubsection{Symbols and Quantities}
\label{sssec:symbandQuant}

% symb forms
The symbolic form framework was developed to explain how successful physics students understand and construct equations \cite{sherin2001}.
In a study of calculus students solving problems in the contexts of areas and volumes, mathematics education researchers observed that students were more productive when they perceived symbols as representing \emph{measures}, including both a value and a unit, throughout their reasoning \cite{Dorko2015CalculusUnits}. This led to the introduction of the \emph{measurement} symbolic form, which includes both a value and an associated unit. The \emph{quantity} symbolic form, introduced in physics education research, modifies this form by including sign as an essential element of a physics quantity \cite{white2019quantification}. The quantity symbolic form represents an important difference between \textit{quantity} in physics and \textit{measurement} in mathematics by including connection to the physical system itself.

%% math symbols
In mathematics education research, literal symbols (letters, sometimes loosely referred to as variables) are classified into several categories including: variables, which represent values that can vary (e.g. $x$, $y$); constants, which represent values that are always the same (e.g. $\pi$); parameters, which represent values that are not changing in that context (e.g. $m$ and $b$ in $y = mx + b$); generalized numbers, which are used in definitions of theorems (e.g. $a$ and $b$ in $ a + b = b + a$), or could represent a quantity (e.g., a block of mass $M$); unknowns, which represent a quantity to be found (e.g. $5x = 13$); and labels, which represent units of measure (e.g. ``m'' for ``meters'') \cite{philipp1992}. Research has demonstrated that literal symbols may invoke student difficulties due to the various roles that they play in a symbolic expression \cite{philipp1992, Trigueros2003}. In mathematics textbooks, problem solving with measures typically involves a measure symbolized by a letter, and often does not include units. The symbols are used throughout the problem; the units are declared at the outset, and tagged on to the solution of the problem. Units are not typically central to mathematical reasoning as part of instruction. 

% physics symbols
Physics, in contrast, often assumes that symbols carry physical information \cite{redish2015}. For example, positive and negative signs play an essential role in physics symbolizing in a way that is sometimes distinct from mathematics, and take on different meanings depending on the physical context \cite{white2020nonip}. Physics symbolizing can be challenging for novices to decode. Prior research demonstrates that physics students are more successful at solving physics problems when physical quantities are given as their numeric values instead of as literal symbols \cite{torigoe2011}. 

Mathematization, proceptual understanding, symbols, and reasoning about quantity are not themselves covariational reasoning; nor are they uniquely important to physics covariational reasoning. They do, however, provide the underpinnings for productive covariation in physics, and play a foundational role for the framework described in the following sections.

\section{Characterizing Covariational Reasoning in Physics}
\label{sec:framework}

In this section, we describe the Covariational Reasoning in Physics (CoRP) framework. We created the CoRP framework by synthesizing the mathematics covariational reasoning frameworks, the recent study of physics experts' covariational reasoning, and other prior work in both mathematics and physics education. The CoRP framework operationalizes the use of covariational reasoning in introductory-level physics, though its application spans the physics curriculum and is not limited to introductory-level physics content. While the CoRP framework is informed by algebra, pre-calculus, and calculus,  this level of mathematics is foundational to quantitative reasoning across the physics major.

The framework consists of three sections: Proceptual Understanding \textbf{(PU)}, Physics Mental Actions \textbf{(PMA)}, and Expert Behaviors \textbf{(EB)}. The Proceptual Understanding section encompasses mathematization and the interplay of quantities and models (symbolic and graphical) in both mathematics and physics contexts. Generally speaking, the proceptual understanding portion of the framework describes mental resources for physics covariational reasoning. \textit{Mathematical Foundations} describes mathematics that is necessary specifically for physics covariational reasoning, and that students may bring with them from previous math courses. \textit{Mathematization in Physics} describes applications of mathematical ideas to physics contexts underlying physics covariational reasoning. 

The Physics Mental Actions are parallel to the Mental Actions presented by mathematics education research (see section IIA and Tab. \ref{tab:mathcovar}) but are attentive to the ways in which recent work suggests that physics experts may reason differently than mathematicians about change and rates of change \cite{expertcovar}. The Physics Mental Actions portion of the framework describes how physics experts think about explicit changes in one quantity, and the effect of those changes on another quantity. 

Finally, the Expert Behavior section describes what experts do when generating models \cite{expertcovar}, using a combination of Proceptual Understanding and the Physics Mental Actions. These behaviors are emergent from prior work \cite{expertcovar}. In this paper, we suggest how they might appear in broader contexts. Some of these behaviors describe how expert physicists reason using the Physics Mental Actions in ways that are distinct from behaviors described in the mathematics education research literature. Others describe behaviors that rely on proceptual understanding of the relevant mathematics and physics content.

Although the framework shown in Table \ref{tab:CRFW} is presented as three distinct parts, there is significant interaction between the three parts when physics experts reason about covarying quantities \cite{expertcovar}. Physics covariational reasoning is complex; the CoRP framework represents one possible operationalization that teases apart foundational reasoning and reasoning about how quantities change with respect to each other.

\begin{table*}[t]
    \centering
    \caption{Current version of a framework to describe the use of covariational reasoning in physics modeling}
    \label{tab:CRFW}
    \begin{tabular*}{0.7\linewidth}{l @{\extracolsep{\fill}}
    l}

         \hline
         \hline
         \multicolumn{2}{c}{\rule{0pt}{1em} \uppercase{Proceptual Understanding}}\\
         \hline
         \multicolumn{1}{l}{\textbf{I. Mathematical Foundations}} & \textbf{II. Mathematization in Physics} \\
         %%%%%%%%%%%%%%%%%%%%%%%%%%%%%%%%%%%%%%%%%%%%%%%%%%%%%%%%%%%%%%%%%%%%%%
         \multicolumn{1}{l}{\quad A. Mathematical Symbols (mathematics)} & \quad A. Mathematical Symbols (physics)\\
         %%%%%%%%%%%%%%%%%%%%%%%%%%%%%%%%%%%%%%%%%%%%%%%%%%%%%%%%%%%%%%%%%%%
         %\multicolumn{1}{l}{\quad1. Symbols} & \quad1. Symbols  \\
         %%%%%%%%%%%%%%%%%%%%%%%%%%%%%%%%%%%%%%%%%%%%%%%%%%%%%%%%%%%%%%%%%%%
         %\multicolumn{1}{l}{\quad2. Role of Variables} & \quad 2. Role of Variables \\
         %%%%%%%%%%%%%%%%%%%%%%%%%%%%%%%%%%%%%%%%%%%%%%%%%%%%%%%%%%%%%%%%%%%
         \multicolumn{1}{l}{\quad B. Common Operations}& \quad B. Constructing Quantities \\
         %%%%%%%%%%%%%%%%%%%%%%%%%%%%%%%%%%%%%%%%%%%%%%%%%%%%%%%%%%%%%%%%%%%
         \multicolumn{1}{l}{\quad C. 7 Parent Functions} & \qquad 1. Mathematical Structure\\
         %%%%%%%%%%%%%%%%%%%%%%%%%%%%%%%%%%%%%%%%%%%%%%%%%%%%%%%%%%%%%%%%%%%
         \multicolumn{1}{l}{}%\quad1. Characteristics of Parent Functions} 
         & \qquad 2. Composite physical quantities \\
         %%%%%%%%%%%%%%%%%%%%%%%%%%%%%%%%%%%%%%%%%%%%%%%%%%%%%%%%%%%%%%%%%%%
         \multicolumn{1}{l}{} & \quad C. Variable Quantities  \\
         \hline
         
       \uppercase{Physics Mental Actions} \hspace{.5em} & \uppercase{Expert Behaviors}  \\
         \hline
         %%%%%%%%%%%%%%%%%%%%%%%%%%%%%%%%%%%%%%%%%%%%%%%%%%%%%%%%%%%%%%%%%%%%%%%
         \textbf{PMA~1 \quad Related Quantities} & \textbf{I. Reasoning Devices}  \\
          %%%%%%%%%%%%%%%%%%%%%%%%%%%%%%%%%%%%%%%%%%%%%%%%%%%%%%%%%%%%%%%%%%%%%%%
         \textbf{PMA~2 \quad Trend of Change} & \quad A. Proxy Quantity  \\
          %%%%%%%%%%%%%%%%%%%%%%%%%%%%%%%%%%%%%%%%%%%%%%%%%%%%%%%%%%%%%%%%%%%%%%%
        \textbf{PMA~3 \quad Coordination of Values} & \quad B. Regions of Consistent Behavior   \\
          %%%%%%%%%%%%%%%%%%%%%%%%%%%%%%%%%%%%%%%%%%%%%%%%%%%%%%%%%%%%%%%%%%%%%%%
     \textbf{PMA~4 \quad Discrete Change} &  \quad C. Physically Significant Points \\
         %%%%%%%%%%%%%%%%%%%%%%%%%%%%%%%%%%%%%%%%%%%%%%%%%%%%%%%%%%%%%%%%%%%%%%%
       \textbf{PMA~5 \quad Small Chunks of Change}&  \quad D. Neighborhood Analysis \\
       %%%%%%%%%%%%%%%%%%%%%%%%%%%%%%%%%%%%%%%%%%%%%%%%%%%%%%%%%%%%%%%%%%%%%%%
       &  \quad E. Compiled Models  \\
       %%%%%%%%%%%%%%%%%%%%%%%%%%%%%%%%%%%%%%%%%%%%%%%%%%%%%%%%%%%%%%%%%%%%%%%
       &  \textbf{II. Modeling Modes}  \\
       %%%%%%%%%%%%%%%%%%%%%%%%%%%%%%%%%%%%%%%%%%%%%%%%%%%%%%%%%%%%%%%%%%%%%%%
       &  \quad A. Function Knowing  \\
       %%%%%%%%%%%%%%%%%%%%%%%%%%%%%%%%%%%%%%%%%%%%%%%%%%%%%%%%%%%%%%%%%%%%%%%
          & \quad B. Function Choosing     \\
       %%%%%%%%%%%%%%%%%%%%%%%%%%%%%%%%%%%%%%%%%%%%%%%%%%%%%%%%%%%%%%%%%%%%%%%
          & \quad C. Symbolic and Graphical Generation   \\  
        % & \textbf{III. Other Tools} \\
        % & A. Assumptions \\
        % & B. Limits \\
        % & C. Symmetry \\
       %%%%%%%%%%%%%%%%%%%%%%%%%%%%%%%%%%%%%%%%%%%%%%%%%%%%%%%%%%%%%%%%%%%%%%%
       
         \hline
         \hline

    \end{tabular*}
\end{table*}

\subsection{Proceptual Understanding}

The Proceptual Understanding portion of the CoRP framework identifies aspects of proceptual understanding of mathematical foundations and mathematization as used in physics covariational reasoning. This section of the framework is divided into math and physics sections to attend to the foundational mathematical reasoning that students bring to physics courses, and the distinct ways that physics uses that mathematical reasoning. Attending to this distinction is an important part of physics instruction \cite{WhiteBrahmia2016}. 

\subsubsection{PU I. Mathematical Foundations}

The Mathematical Foundations portion of the CoRP framework describes the aspects of mathematical reasoning that are necessary for productive covariational reasoning in physics. Generally speaking, introductory physics students see and learn these ideas from prerequisite math courses, though they may not display the same fluency as physics experts.

\textit{A. 
%Mathematical Foundations:
Mathematical Symbols} is characterized by using symbols to represent mathematical concepts and \textit{measures} (i.e., a value and an associated unit, as described by the measurement symbolic form \cite{Dorko2015CalculusUnits}). This framework element includes, but is not limited to, symbols that represent values (e.g., $x$, $\pi$) and operations (e.g., $+$, $-$) as used in mathematics.

\textit{B. Mathematical Foundations: Common Operations } is characterized by the use of operations that are ubiquitous in introductory physics and calculus classrooms, including but not limited to addition, multiplication, subtraction, division, taking a derivative or limit, and integration. Making sense of the meaning and contextual relevance of these operations is essential for covariational reasoning in physics.

\textit{C. Parent Functions } is characterized by reasoning about a handful of common functions, chosen because they are the most common functions used in introductory physics: 
\begin{itemize}
    \item linear ($y \propto x$), 
    \item quadratic ($y \propto x^2$), 
    \item sine / cosine ($y \propto \sin{x}$), 
    \item inverse ($y \propto 1/x$), 
    \item inverse square ($y \propto 1/x^2$), 
    \item exponential ($y \propto e^x$), and 
    \item logarithmic ($y \propto \ln(x/x_0)$). 
\end{itemize}
We adopt the mathematics language of ``parent functions,'' commonly used when teaching functional transformations, to illustrate that we are referring to the functional relationship between the variables. Essential background knowledge involves a general familiarity with the behavior of parent functions, including:

\begin{itemize}
    \item the ability to sketch a graph of a given parent function, or associate a graph with a parent function,
    \item the ability to describe the general behavior of the function, including concavity and end behavior. 
    \item familiarity with function transformations (stretching, translation, etc.) in order to use them in a wide variety of scenarios. 
\end{itemize}

\subsubsection{PU II. Mathematization in Physics}

Mathematization in Physics describes the foundational quantitative reasoning about physics quantities that is necessary for productive physics covariational reasoning. Introductory physics students may not have ample experience with mathematization from prior coursework. 

\textit{A. 
%Mathematization in Physics: 
Mathematical Symbols} is characterized by symbolizing values and physical constants (e.g., $G$, $\pi$), operations  and physical quantities which might be a variable, parameter, or general variable in a particular context. This also includes recognition, based on context, of a symbol as a representation of a given quantity.

\textit{B. Constructing Quantities} is characterized by using common operations to construct a quantity, or to make sense of how a quantity is constructed. Here, we focus on three aspects of constructing quantities.

\begin{itemize}
    \item[1.] Reasoning about the \emph{mathematical structure} of a quantity includes recognition of features of the representation of the quantity---for example, whether the quantity includes a direction or a sign. This category includes recognizing physical attributes of a given quantity (e.g., whether it can be positive or negative, or discrete or continuous) and understanding how those attributes will be represented symbolically or graphically. 
    \item[2.] Combining two or more quantities to create a new, \emph{composite physical quantity} is ubiquitous in physics. Most physics quantities are product or ratio combinations of the seven base quantities (length, time, amount of a substance, electric current, temperature, luminous intensity, mass). Graphical features such as slope (a ratio) and area under the curve (an accumulated, multiplicative quantity) are important composite physical quantities. 
\end{itemize}

\textit{C. Variable Quantities} is characterized by a recognition of which quantities in an equation make sense to vary, how they vary, and which other quantities don't vary. Physical models typically involve many symbols in which the literal symbols might represent constants, general variables, parameters, or varying quantities; the classification of a particular literal symbol can change from context to context---sometimes even within the same problem. A proceptual view of variable quantities includes being able to reason about which quantities are varying with respect to one another, and which represent parameters or constants.Variable Quantities also includes paying attention to the units of a function. For example, since a graph is a representation of how the dependent variable changes relative to changes in the independent variable, the points on the curve take on the units of the dependent variable, and the slope is a rate of change, which has units of the dependent variable over the independent variable.

\subsection{Physics Mental Actions}

Physics Mental Actions (PMA) describe the explicit consideration of the change in one quantity as the result of the change in another quantity. Though the PMA are similar to the mental actions described by mathematics education researchers (See Tab. \ref{tab:mathcovar}), they differ in a key way: a focus on \emph{quantities} as the objects of covariation. This results in not only a superficial change (i.e., using the word ``quantity'' rather than the word ``variable'' in the descriptions) but also modifications that make them more consistent with the ways that physics experts use them with physics quantities. Expert-like reasoning about a quantitative relationship between quantities is generally not separable from reasoning about the physics quantities themselves \cite{czocher2017can,strawberryfields}. The Physics Mental Actions are often guided by what is physically reasonable \cite{expertcovar}. For example, physics experts often rely on understanding of how a quantity can change (e.g., continuously or discontinuously), or whether a change in one quantity would, in the real world, cause the change in another. The PMA are not hierarchical in the sense that PMA~5 is ``better'' than PMA~1; however, with the exception of PMA~3, the PMA are listed in order of increasing specificity about how quantities are related. PMA~3 is not included in this hierarchy because it does not include reasoning about change; we include it in the framework nevertheless because it describes a way that physicists relate quantities.

In the covariational reasoning frameworks developed by mathematics education researchers, only one mental action (MA~4) is associated with consideration of discrete change \cite{carlson2002,thompson2017,Jones2022MultivariationReasoning}. All instances of discrete covariation are associated with MA~4, regardless of the size of the discrete ``chunk.'' Different considerations of how quantities change in discrete chunks led to development of two PMA related to discrete covariational reasoning \cite{Ely2019}. PMA~4 is most similar to what physics education research has termed scaling \cite{Arons1976, Trowbridge1981, Bissell2022}, and typically involves large, often integer-valued chunks. Prior work suggests that physics experts rarely if ever consider smooth, continuous changes of multiple quantities simultaneously \cite{expertcovar}, as described by mathematics education researchers' MA~5; instead, physics experts engaging in PMA~5 are likely to consider small ``chunks'' of change (e.g., considering $dx$ to be a very small $\Delta x$) \cite{expertcovar,kustusch2014partialAnalysis}. For this reason, we do not include a PMA that is analogous to MA~5, but think of PMA~5 as the ``most continuous'' instance of discrete covariation.

\noindent\textbf{PMA~1: Related Quantities} 

PMA~1 is characterized by a recognition that one quantity is related to another quantity. Some related behaviors include labeling axes of a graph, and a verbal acknowledgement that if one quantity changes the other will as well. This typically includes a choice of which quantity is the independent quantity and which is the dependent quantity, often guided by understanding a cause and effect relationship between the quantities. The recognition that the potential energy of a spring changes as a result stretching the spring from its equilibrium length is an example of PMA~1.

\noindent\textbf{PMA~2: Trend of Change} 

PMA~2 is characterized by describing whether a quantity will increase or decrease as a result of another quantity increasing or decreasing. Some related behaviors include drawing arrows to indicate increases or decreases, drawing graphs that represent linear approximations, and verbalizing the trend of change. The recognition that the potential energy of a spring increases as the spring is stretched is an example of PMA~2.

\noindent\textbf{PMA~3: Coordination of Values} 

PMA~3 involves tracking the values of two quantities to create a discrete set of associated pairs. Related behaviors involve plotting points or creating a table of values. This does not necessarily entail consideration of simultaneous change of both quantities, but rather considering multiple values of one quantity and determining the associated values of another quantity for those values. Determining the value of the potential energy of a spring for a discrete set of values of the amount that a spring is stretched from its equilibrium length is an example of PMA~3.

\noindent\textbf{PMA~4: Discrete Change} 

PMA~4 refers to reasoning around what happens to one quantity if another, related quantity changes by a substantial, discrete amount. It is most often characterized by what is sometimes called ``scaling'' in physics education research: considering how changing one quantity by a multiplicative factor affects another quantity (e.g., ``if I triple x, what happens to y?''). In some instances, it is consideration of change by a substantial, additive amount. Some related behaviors include plugging in numbers and comparing the change, considering how the dependent quantity changes with a substantial change in the dependent quantity, and verbalizing multiplicative changes. Recognizing that doubling the stretch of a spring results in the spring potential energy increasing by a factor of four is an example of PMA~4.

\noindent\textbf{PMA~5: Small Chunks of Change} 

PMA~5 is characterized by reasoning about the resulting change in one quantity due to small, discrete changes made to another, related quantity. This reasoning is grounded in examining what happens for \emph{small} pieces of change. Some related behaviors include ``zooming in'' to a graph by examining the slope for a small region \cite{Ely2019}, verbalizing an awareness that a change is small compared to the scale of the problem, and moving fluidly between representations of discrete change and derivative notation \cite{kustusch2014partialAnalysis,dray2010putting}. Recognizing that the potential energy of a spring changes slowly near the equilibrium point, and more quickly further from the equilibrium point is an example of PMA~5.

\subsection{Expert Behaviors}

In this section, we describe a number of specific physics covariational reasoning behaviors that we call \textit{Expert Behaviors}. The Expert Behaviors portion of the CoRP framework is not independent of either Proceptual Understanding or the Physics Mental Actions; rather, expert behaviors are dependent on a proceptual understanding of the underlying mathematics and the relevant physics quantities, and use of the Physics Mental Actions. 

The expert behaviors described below are based on reasoning and approaches seen in physics experts' reasoning while completing graphing tasks designed to elicit covariational reasoning \cite{expertcovar}.

\subsubsection{EB I. Reasoning Devices}

A reasoning device is a tool or small piece of reasoning that is employed while reasoning covariationally  \cite{expertcovar}. Some of the reasoning devices are associated with the Physics Mental Actions. In this section, we define each of the CoRP framework reasoning devices. Then we give an example of how the reasoning devices could be used together to create a graphical representation using a context typical of an introductory physics back-of-chapter problem.

\textit{A. Proxy Quantity} describes when a quantity is substituted for another while covarying two quantities. 
Use of a proxy quantity may allow for easier covariational reasoning. Use of proxy quantities is fundamentally related to related quantities (PMA~1), the recognition that two quantities are related. 

A proxy quantity may be used to make a novel physics context into a more familiar context. For example, for constant motion contexts, time is commonly used as a proxy quantity when the task specifies ``total distance traveled'' as one of the quantities to be covaried \cite{expertcovar}. Considering how a quantity changes with respect to time is oft-practiced in physics; that is, time is a very frequently the independent quantity involved when related quantities (PMA~1) is engaged. Using time as a proxy quantity in this way often allows for easier, more rapid covariational reasoning, or the application of a familiar model. Use of time as a proxy quantity is consistent with work by mathematics and chemistry education researchers \cite{johnson2017ferris,Johnny2020}.

A proxy quantity may also be used because of how particular quantities are understood. For example, electric potential is typically understood as a function of position $r$ as measured relative to the source (i.e., $V = V(r)$). When asked to covary potential $V$ with a quantity other than $r$, it is more familiar to use $r$ as a proxy quantity \cite{expertcovar}.  This use of proxy quantity differs from the use of time as a proxy quantity in one important way: $r$ is not necessarily directly proportional to the quantity that it substitutes. This, again, engages PMA~1, as some quantities are considered to be functions of another quantity, and are therefore necessarily related.

\textit{B. Region of Consistent Behavior} describes when a domain is separated into sections that could be modeled by the same function, or where the behavior of the relevant system is constant or consistent in some way. An expert may break up a domain into regions of consistent behavior in order to associate a single trend, function, or model with each section as appropriate, and may consider each region separately when constructing a graphical or symbolic representation. Sometimes, regions are determined by whether a quantity is increasing, decreasing, or constant; in these cases, this behavior is related to trend of change (PMA~2). For example, a graphical representation of a car speeding up steadily, then moving at a steady speed, and finally slowing down steadily between two stoplights would be represented by three distinct regions on a graph of $v$ vs $t$. 

\textit{C. Physically Significant Points} describes the identification of a point that holds physical meaning for a given context. An expert may choose and plot a small number of physically significant points to begin to construct a model. Examples of physically significant points are: a boundary between regions of consistent behavior, a bound of a quantity (a local or absolute maximum or minimum), or a point where a quantity changes the most or least rapidly. These points can be used to guide the construction of either a graphical or symbolic representation. Identification of physically significant points is an example of coordination of values (PMA~3).

\textit{D. Neighborhood Analysis} involves use of the Physics Mental Actions around physically significant points. This device is typically used in the construction of a graphical representation by considering the rate of change of the quantities around chosen points, and drawing a small line segment centered on the point to indicate the slope of the graph at that point. In other cases, neighborhood analysis can be used to check the appropriateness of a constructed representation. Experts may consider whether a constructed graphical representation models the behavior of the quantities correctly near physically significant points. This reasoning could be extended to the construction and checking of a symbolic representation by considering how the derivative of an expression should behavior at certain values; for example, whether the derivative should be positive, negative, or zero at a point. Neighborhood analysis involves reasoning about small chunks of change (PMA~5), as it involves covariation of quantities for small deviations near a physically significant point.

\textit{E. Compiled Models }are usually rapid, almost-automatic associations between a relevant parent function and a given physics context. Experience with physics problem-solving often includes having an almost-automatic association between a physics context and a function \cite{expertcovar}. For example, a physics expert may associate: circular motion with a sinusoidal function; the trajectory of a projectile with a quadratic function; or the electric potential near a point charge with the function $1/r$. Use of a compiled model requires trivial use of PMA~1 (a recognition that the quantities are related), but typically does not rely on other Physics Mental Actions. 

\begin{figure}
    \begin{subfigure}[b]{\linewidth}
    \centering
        \includegraphics[width=\linewidth]{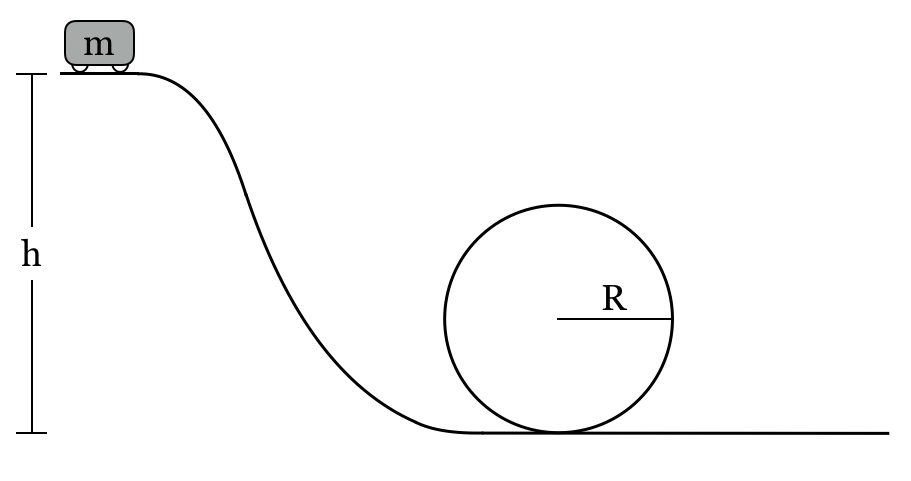}
        \caption{Task figure. The loop portion of the track is estimated as circular.}
    \end{subfigure}
    \begin{subfigure}[b]{\linewidth}
    \centering
        \includegraphics[width=\linewidth]{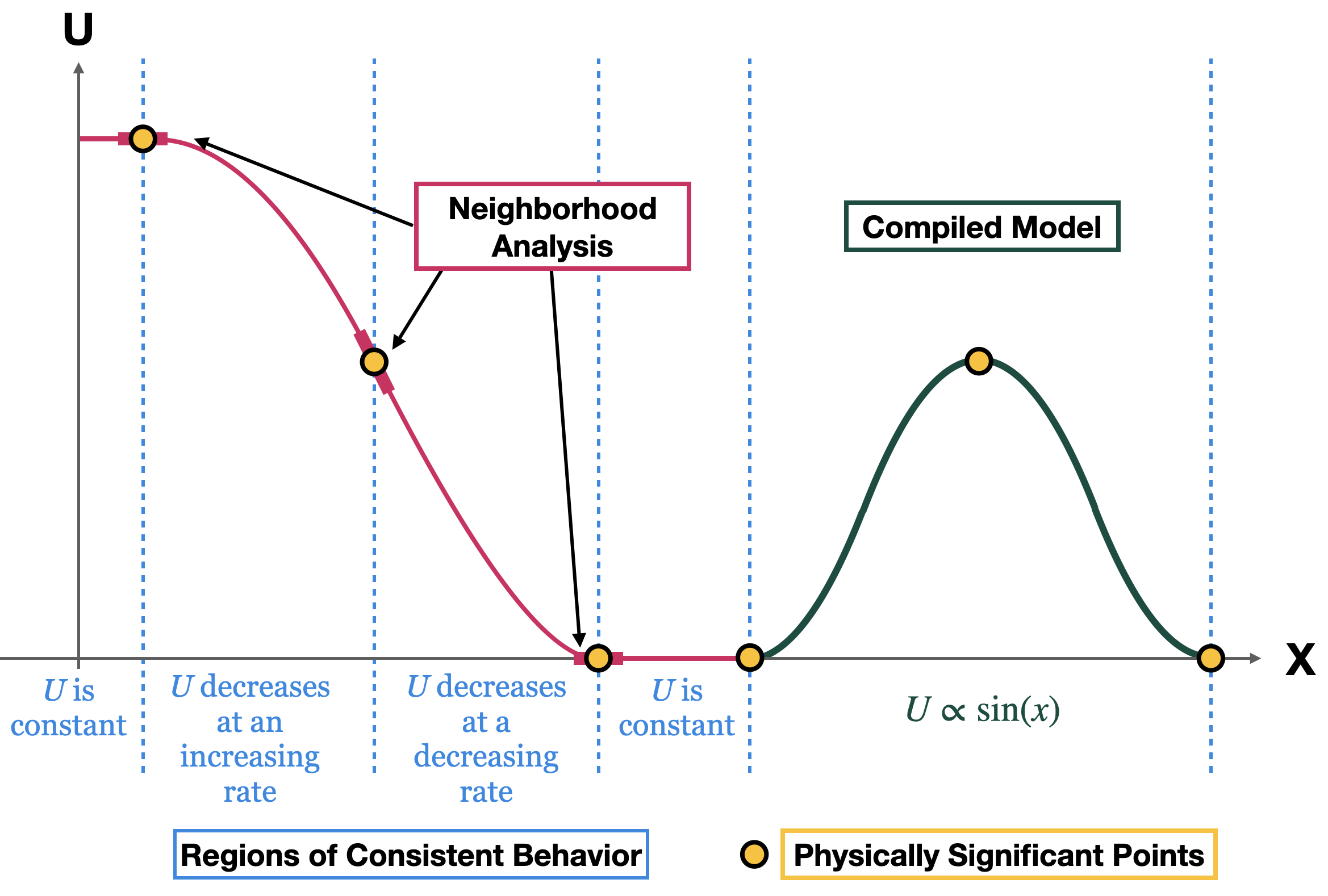}
        \caption{Annotated qualitatively correct graph for task. }
    \end{subfigure}
    \raggedleft
    \caption{Task and example graph provided here to exemplify use of the Expert Behavior reasoning devices. For this task, create a graph of the potential energy of the car-Earth system as a function of distance traveled for a toy car of mass $m$. }
    \label{fig:toyCar}
\end{figure}

\textbf{Example Application:} To show how the reasoning devices can be used to create a graphical representation, we use the example shown in Fig. \ref{fig:toyCar}. The task asks for a graph of the gravitational potential energy of the car-Earth system for a toy car on a track 

While this specific question has not been used as an interview task with physics experts, the tasks in the related study are quite similar in structure \cite{expertcovar}. Thus, we are not claiming that the following is evidence of expert reasoning. We present what follows to exemplify use of the reasoning devices (EB~IA--E) as was observed with similar tasks \cite{expertcovar}. 

To begin this task, one could recognize that the potential energy of the system is directly proportional to the height of the car. Therefore, the height of the car can be used as a proxy quantity (EB IA) for gravitational potential energy. Next, one could break the motion of the task into several regions of consistent behavior (EB IB): the flat portion at the top of the hill, the hill portion, the flat portion of the track between the hill and loop, and the circular loop. This could be done in tandem with the identification and plotting of physically significant points (EB IC): the potential energy is at a maximum at the beginning of the hill, when the car has traveled zero distance; the potential energy is at a minimum at the end of the hill, when the car is about half-way through its journey. Similarly, one might note the beginning of loop, top of loop, and end of loop (see Fig. \ref{fig:toyCar}). 

One might then reason about what happens between the plotted points. During the hill portion, one may use neighborhood analysis (EB ID) to identify that the potential energy is changing most rapidly toward the middle of the hill, and indicate that with a steep, negative line segment about halfway between the points at the top and bottom of the hill. For the circular section, one may recognize that the height is related to time (proxy quantity, EB IA), and apply a compiled model (EB IE) in which the height varies sinusoidally with time for circular motion contexts. Finally, an expert may recognize that the height changes smoothly, and connect the points with a smooth curve. \\

\subsubsection{EB II. Modeling Modes} 

In addition to the reasoning devices described above, we include approaches to modeling novel scenarios that have been observed in physics experts engaged in covariational reasoning tasks, referred to as \textit{Modeling Modes} \cite{expertcovar}. Modeling Modes describe the ways that experts may construct a specific mathematical representation for a given physics context. For the purposes of this work, we define a mathematical representation as a symbolic expression, or a graph with or without an accompanying symbolic expression \footnote{We do not attempt to describe construction of covariational diagrammatic representations with this framework, though we recognize their importance.}. The Modeling Modes involve proceptual understanding of the relevant mathematical foundations and physics quantities, Physics Mental Actions, and Reasoning Devices. In the descriptions below, we use some examples that are familiar to the reader to help illustrate the reasoning modes.

\textit{A. Function Knowing} is a behavior that relies on PMA~1 (the recognition that two given quantities are related to each other), use of a Compiled Model (EB~IE) that associates a parent function with a given context, and proceptual understanding of the relevant parent function (PU~IC) itself to associate a quantitative representation to a given physics context.

Function Knowing can be used to create a graphical representation by considering relevant parameters, plotting a few points based on those parameters, and connecting those points with the known function (e.g., for an object in circular motion at a steady speed, the initial conditions such as starting position and radius of the circle can be used with a familiar model for uniform circular motion). Relevant parameters can be similarly used to create a symbolic representation for the specific context. This type of ``knowing'' is typically guided by the physics content and draws on well-tested models of nature that are familiar to the expert. 

Function Knowing can be accessed in a wide range of physics contexts, based on well-established connections between the context and a function-based model, such as the connection between uniform circular motion and sine or cosine functions; potential near a point source and $1/r$; as well as many, many contexts with direct proportionality.

\textit{B. Function Choosing} is the behavior of using a combination of physically significant points (EB~IC), trend of change (PMA~2) and compiled models (EB~IE) to select one of several possible functions that might be fruitful in a particular context. This process is informed by physics content knowledge of the context, and a proceptual understanding of parent functions.

For example, to generate a model of the relationship between the gravitational potential energy of a system of two objects and the distance separating the objects, one may initially identify a trend (PMA~2)---that is, that gravitational potential energy decreases as the distance between the objects decreases. Lacking additional information, assumptions about the context guide the choice of a compiled model (EB~IE): assuming that the system consists of a relatively small object near the surface of a very massive planet may lead to the assumption that the change in potential energy is linear, whereas farther from the surface of the planet, the potential energy could be assumed to decrease as $1/r$. To create a graph, one may define symbolic values for initial and final positions and potential energies (EB~IC), plot those points, and connect them with the function chosen based on any assumptions. Similarly, to create a symbolic expression, one could symbolize the relevant quantities, and then use them with the chosen compiled model (EB~IE)---in this case, an association between the expressions $U = mgh$ and $U = -\frac{GMm}{r}$ with gravitational potential energy.

\textit{C. Symbolic and Graphical Generation} refers to behaviors that use consideration of covarying quantities to generate an explicit covariational relationship---either with a generalized symbolic representation or a qualitatively correct graph. If a graph is generated, it may (or may not) be used to identify an appropriate parent function that can be used as the basis of a refined symbolic expression.

Symbolic generation is based on prior research in physics education. Sherin's work on symbolic forms focuses on how physics students interpret and create symbolic expressions \cite{sherin2001}. In that work, he describes how students might use small pieces of reasoning about change to develop a generalized quantitative model. For example, one student is reported to state: ``The coefficient of friction has two components. One that's a constant and one that varies inversely as the weight.'' The student then develops the expression: $\mu = \mu_1 + C \dfrac{\mu_2}{m}$ \cite{sherin2001}. Notably, the product of symbolic generation is not necessarily a fully developed quantitative model. Rather, it is characterized by using symbols to express a covariational relationship between quantities.

Symbolic and graphical generation typically occurs when there is no known quantitative model for a context \cite{expertcovar}. One approach begins with the identification of physically significant points (EB~IC) and neighborhood analysis (EB~ID), which involves consideration of how quantities covary around the physically significant points (PMA~5). Small line segments representing the approximate rate of change of the height with respect to time around those points can be drawn and then connected together in ways deemed appropriate for the context.

\begin{figure}
    \centering
    \includegraphics[width=0.7\linewidth]{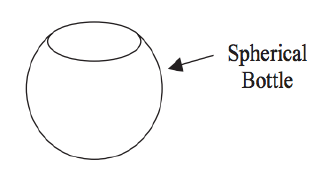}
    \includegraphics[width=0.7\linewidth]{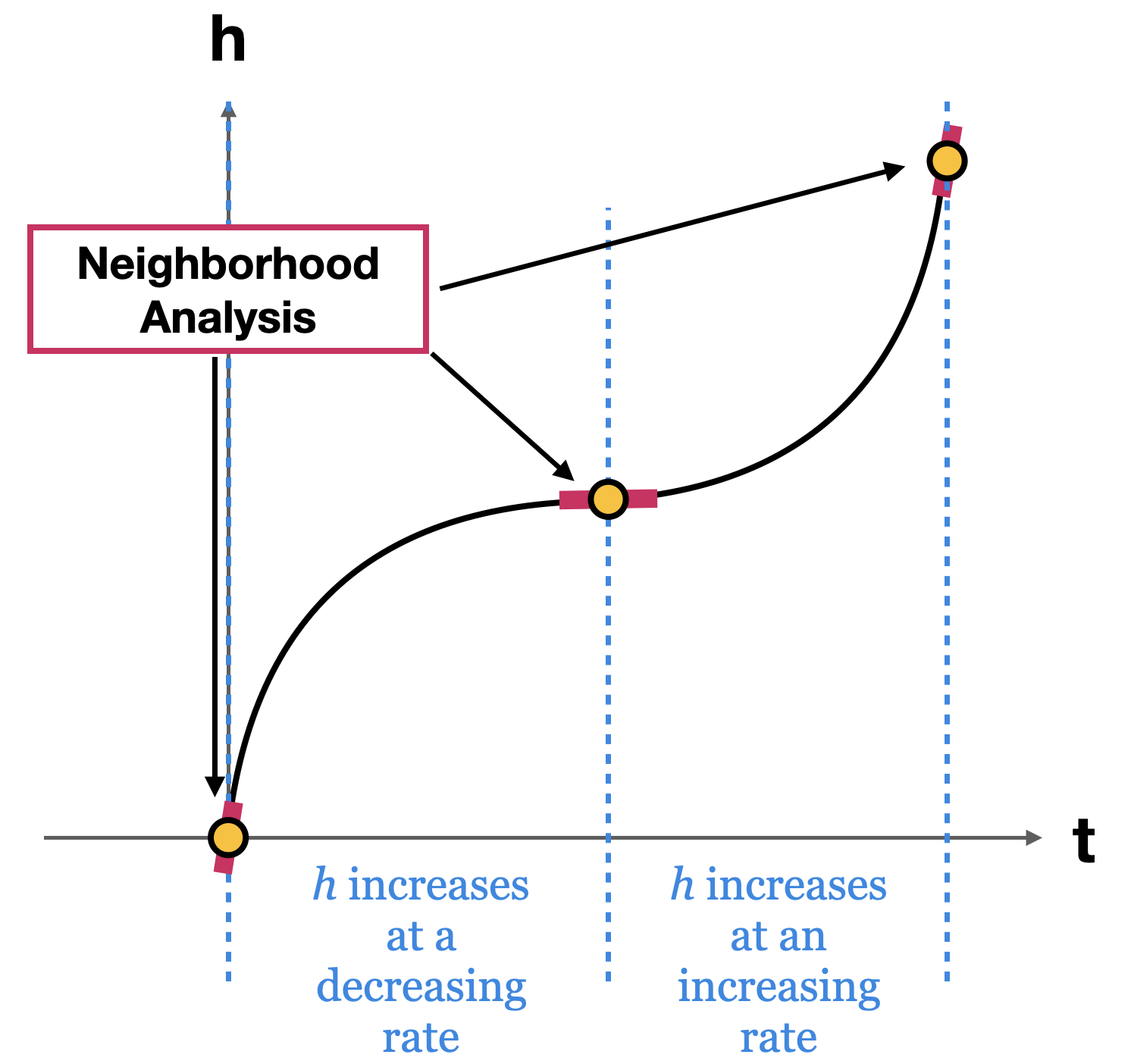}
    \caption{Top: Bottle-filling task figure \cite{carlson2002}. Bottom: A solution to the bottle-filling task, showing expert behaviors that may be used.}
    \label{fig:sphericalbottle}
\end{figure}

Consider the \textit{Spherical Bottle} task shown in Fig. \ref{fig:sphericalbottle}, in which water is poured at a constant rate into a spherical bottle of radius $R$.  Variations of the bottle-filling task have been used to assess students' covariational reasoning \cite{carlson2002,johnson2013connecting}. To create a graph of how the height $h$ of the water in the bottle varies with time, one could consider three physically significant points (EB~IC): a point near the base of the bottle, a point at the middle of the bottle, and a point near the top of the bottle. The height of the water at these points will be $0$, $~R$, and $~2R$, respectively. After plotting these points, neighborhood analysis (EB~ID) with covariation of height and time (PMA~5) around them can be done. At the top and bottom, where the bottle is more narrow, the height will change quickly. At the middle, where the bottle is wider, the height will change more slowly. Based on this analysis, small line segments representing the rate of change of the height with time can be added around the identified points. The line segments should be steeper at the points at $h=0$ and $h=2R$ and less steep at $h=R$. Given that the height should be changing smoothly, with no discontinuities or sharp points, the line segments can be connected with a smooth curve. 

An expert may (or may not) identify that the created graph looks like a cubic function, and create a symbolic expression that represents the graph.

\section{Relevance of the covariational reasoning framework}
\label{sec:CoRPrelevance}

In this section we demonstrate how the CoRP framework can be used to analyze student reasoning. In particular, the framework allows us to identify reasoning resources that students have about covariation at various levels of instruction, allowing for a glimpse into reasoning that might be in their \textit{zone of proximal development}---the space between what a learner can do without assistance and what a learner can do with instruction or collaborative help \cite{Vygotsky1978}---at each stage. The analysis of students at various levels of physics can contribute to characterizing the novice-to-expert continuum.

We begin with analyses of math students' covariational reasoning, as reported in mathematics education literature. These analyses are useful because many students develop covariational reasoning in math courses first, before coming to physics. As such, we suggest that algebra and calculus students are representative of students coming into introductory physics, and analysis of their covariational reasoning through the lens of the CoRP framework can help us identify the resources they bring with them from mathematics courses. These analyses also indicate opportunities for targeted instruction to develop physics students' physics covariational reasoning. Next, we analyze studies from the current literature of physics students at both introductory and upper-division undergraduate levels, to highlight some of the physics covariational reasoning used across the major. By looking at the reasoning of students through the lens of the CoRP framework, starting with students enrolled in math courses and progressing to upper-division physics students, we can characterize some trends in physics covariational reasoning development with progressively more physics instruction.

\subsection{Identifying resources of algebra and calculus students}
Mathematics education researchers Johnson, McClintock, and Hornbein report on a case study of high school algebra student ``Ana'' \cite{johnson2017ferris}, who participates in a number of interviews about various covariational reasoning tasks. For a variation of a bottle-filling task (similar to the task shown in Fig. \ref{fig:sphericalbottle}, above), Ana is shown an animation of water being poured into a spherical bottle with a cylindrical neck at the top, and is asked to create a graph to relate the amount of water in the bottle and the height of the water in the bottle. Initially, Ana creates a linear graph, saying ``the height would still increase\ldots and the volume would still increase.'' This is consistent with identifying a trend of change (PMA~2), as Ana is recognizing that the height increases as the volume increases. After some probing from the interviewer, Ana creates a piecewise linear graph, saying,

\begin{quote}
\ldots the water is filling up very fast, and then it's like slower, and then it's really fast. Really fast would happen here [indicating neck of bottle], and kind of fast here [indicating bottom of bottle], and slow here [indicating middle of spherical portion of bottle].
\end{quote}

Through the lens of the CoRP framework, we characterize this statement as proto-expert-like: Ana is beginning to identify Regions of Consistent Behavior (EB~IB), and is doing an early version of Neighborhood Analysis (EB~ID) by considering differences in the rates at which the height changes in different regions in the bottle. However, rather than considering ``instantaneous'' rates of change around physically significant points (EB~IC), Ana seems to consider average rates over larger regions. This is consistent with use of larger chunks of change (PMA~4), rather than the small chunks of change (PMA~5) that experts typically consider for neighborhood analysis.  According to the sections of transcript provided by the researchers, Ana does not consider what the physical significance of a linear (or piecewise linear) graph would be and does not generally consider the meaning of the slope of the graphs she constructs. 

Mathematics education researchers Carlson, Jacobs, Coe, Larsen, and Hsu report on 20 college-level calculus  students' performance on the same variation of the bottle-filling task (a spherical bottle with cylindrical neck) \cite{carlson2002}. Carlson et al. report that only one student (``Student D'') created a linear graph, and the student justifies their graph saying ``as the volume comes up, the height would go up at a steady rate.'' While it is tempting to classify this reasoning as indicative of PMA~2 (Trend of Change), we believe that because the student makes a connection between the linearity of their graph and the identified ``steady'' rate that the reasoning is more consistent with application of a compiled model (EB~IE) or function knowing (EB~IIA). A compiled model (EB~IE) is not associated with the Physics Mental Actions---that is, there is rarely explicit consideration of how a change in one quantity affects another quantity. Rather, use of a compiled model is typically built on familiarity with and understanding of a given context and function. In this case, use of a straight line to indicate a constant rate implies understanding of why a constant slope is associated with a constant rate. We note that a linear functions are often a default assumption for unfamiliar contexts for both students and experts \cite{expertcovar,VanDenEynde2019,strawberryfields}. 

The remainder of the students in the study by Carlson et al. created non-linear graphs, suggesting an awareness that the height of the water is not directly proportional to the amount of water in the bottle. A majority of the remaining students created graphs that were increasing and entirely concave-up or concave-down. ``Student B'' explains their concave-down graph by saying,

\begin{quote}
    Okay, the more water, the higher the height would be \ldots Right here [indicates bottom of bottle] the height will be zero and the volume is zero. As you go up, a little more height increases and the volume increases quite a bit, so the amount by which the height goes up is not as much. Once you get there [indicates halfway up the spherical part of the bottle], the height increases even slower\ldots So, every time you have to put more and more volume in to get a greater height towards the middle of the bottle and once you get here [pointing to the top of the spherical portion], it would be linear, probably. So, it's always going up [tracing his finger along the concave-down graph], then it would be a line.
\end{quote}

Student B begins by indicating a trend of change (PMA~2) and by coordinating values (PMA~3) at a physically significant point. The student continues by engaging in a preliminary version of neighborhood analysis (EB~ID). Expert-like neighborhood analysis involves identification of physically significant points, and use of PMA~5 in small regions around those points; here, Student B instead considers regions, starting at the bottom of the bottle and moving towards the top, and compares relative rates of change in those regions. Unlike Ana, Student B does seem to distinguish between average and instantaneous rates of change. Student B associates the shape of their graph with a perceived decreasing rate of change, having to add ``more and more volume'' for a given change in height. This is suggestive of reasoning about small chunks of change (PMA~5). When reasoning about the cylindrical neck of the bottle, Student B's statement that ``it would be linear, probably,'' is consistent with a compiled model relating cylinders and volume proportional to height. 

From this review of math students' covariational reasoning, we identify facility with linear functions as a reliable resource. All of the math students discussed here are able to reason using the spectrum of Physics Mental Actions for linear relationships. 

The calculus students displayed additional emerging resources that are important for covariational reasoning in physics: 

\begin{enumerate}
    \item Reasoning about changing rates of change. While we don't consider reasoning about rates of change for linear functions to be less important than non-linear functions, we do recognize that reasoning about non-constant rates of change is more challenging for students in courses in both mathematics and physics. 
    \item Interpretation of the slope of a constructed graph as a rate of change, and association of a linear function with a physical context. Both of these lines of reasoning indicate sensemaking about a physical context with mathematics. 
\end{enumerate}

\subsection{Identifying resources of physics students}

In this section, we analyze prior work in physics education research through the lens of the CoRP framework. Development of instruction targeted at improving students' physics covariational reasoning requires understanding what types of reasoning are in students' zones of proximal development. Little work has been done in physics education research to study physics students' physics covariational reasoning, and how it may progress with instruction. Therefore, recognizing changes in physics covariational reasoning as a result of physics instruction is more difficult than characterizing the development of covariational reasoning by analyzing math students' reasoning. However, the CoRP framework provides a lens for understanding the existing work in a new light. These analyses also suggest that a focus on students' physics covariational reasoning could be a productive area for future research.

In physics, where quantities are often not easily visualized, making sense of the meaning of quantities and their covariational relationships becomes more difficult---and more important to learning. Sherin's symbolic forms framework describes how physics students understand and construct physics equations \cite{sherin2001}. Observations of advanced-introductory physics students informed Sherin's development of the symbolic forms framework, which describes multiple aspects of how students reason about symbolic expressions. Here, we look at some of the students described by Sherin to begin to characterize how introductory physics students reason covariationally when engaged in quantitative modeling using symbolic expressions. We note that Sherin emphasizes in his paper that students often needed prompting from the interviewer (a physics expert) in order to display some of the behaviors that he later categorized as symbolic forms. 

In one task, Sherin asked students to determine how a dropped object's terminal velocity depended on its mass. In working on this task, all of the students recognized that the force of air resistance is dependent on the velocity of the object (PMA~1), and that the air resistance increases as the velocity of the object increases (PMA~2). Sherin reports that at least one student reasoned mechanistically about why this must be the case, saying that the force of air resistance ``gets greater as the velocity increases because it's hitting more atoms \ldots of air.'' Some students then engaged in an incomplete version of function choosing (EB~IIB), trying to decide the functional relationship between velocity $v$ and the force of air resistance $F_U$. One student, ``Jack,'' listed a number of possibilities, writing
\begin{quote}
\begin{equation*}
    \begin{aligned}
        F_u &\propto v\\
            &\propto v^2\\
            &\propto v^3\\
    \end{aligned}
\end{equation*}
    
\end{quote}

and then saying:

\begin{quote}
    Somehow they're related to the velocity but we're not sure what that relationship is. All we can say is that $F_U$ is either proportional to $V$ or to $V$ squared or to $V$ cubed.
\end{quote}

While Jack is able to generate a number of possibilities, he is not able to choose one as being the most appropriate for the situation, nor does he state why those are the only possibilities. Using assumptions, facility with common operations (PU~IB) and parent functions (PU~IC), and familiarity with the physics content to choose an appropriate function for a given context are key components of function choosing. Jack (and others) did not eliminate functions or choose one option as being more reasonable than others. This suggests that function choosing is an emerging, rather than reliable, behavior for these students.

Similarly, for a task involving a spring, students ``Mike'' and ``Karl'' struggled to remember whether the expression for the force exerted by a spring was $F = k x$ or $F=\frac{1}{2}kx^2$. 

\begin{quote}
    Karl: Okay, now, qualitatively, both $kx$ and half $kx$ squared do come out to be the same answer because as [pause] looking at it qualitatively, both half---both half $kx$ squared and $kx$, um, you know, increase as $x$ increases. 
\end{quote}

Mike and Karl recognized that the force exerted by the spring depended on the stretching of the spring (PMA~1) and that it would increase as the spring was stretched (PMA~2), but could not use other information to help them decide which was more appropriate. We note, for example, that the students couldn't rely on  solid understanding of force as a quantity (PU~IIB) to determine which equation could be correct; $\frac{1}{2}kx^2$, the expression for spring potential energy, is necessarily a scalar quantity and could not be the correct expression for a force, a vector quantity.

\begin{figure}
    \centering
    \includegraphics[width=0.49\textwidth]{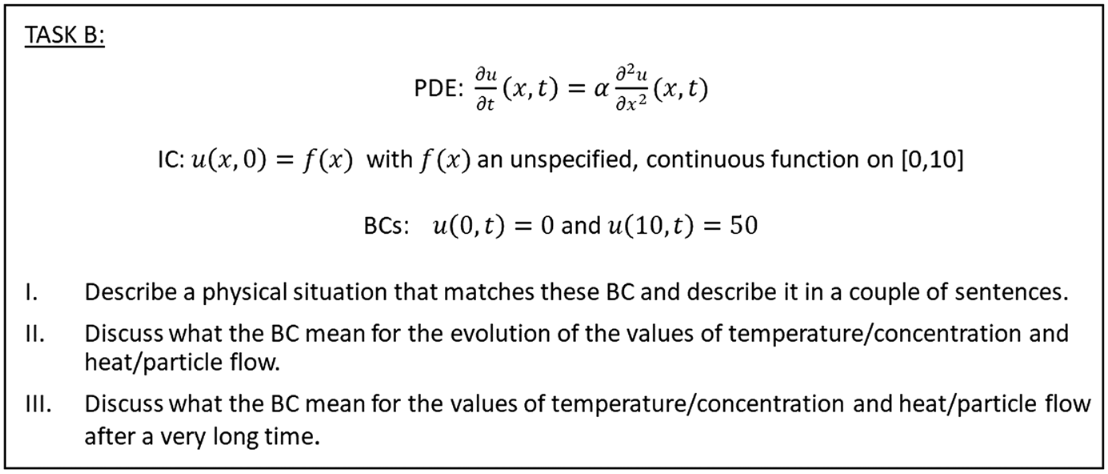}
    \caption{Graphing task used with upper-division physics students in study by Van den Eynde, et al. The students are working on II.}
    \label{fig:heatFlowTask}
\end{figure}

A proceptual understanding of quantity is valuable and relevant for contexts involving more advanced physics. Physics education researchers Van den Eynde, Schermerhorn, Deprez, Goedhart, Thompson, and De Cock describe a case study of second-year physics and math students reasoning covariationally in the context of the heat equation \cite{VanDenEynde2019}. In the Van den Eynde et al. study, students were prompted to generate graphs of the relationship between heat flow and time based on the information in the task (see Fig. \ref{fig:heatFlowTask}). 

One pair of students,``Evan'' and ``David,'' appear to engage in function choosing (EB IIB):

\begin{quote}
    Evan: It [the graph] flattens out again, because that, that is just what things do under influence of the heat equation, but how would it look like at the end? Something like, eh, straight line or so maybe? I don't really know\ldots What makes most sense?\ldots
    
    David: Wait, the end should be kept at 50, so there should be some kind of heat source.

    Evan: Yes.
    
    David: Yes, and if there is a heat source, heat is again flowing over there, to the beginning of the
    rod (referring to the boundary condition at position 0). But the beginning of the rod is so cold that it stays zero and so it cancels each other out a bit and I think you will just get a straight line (10).

    Evan: Yes, I also have the feeling it will become a straight line, because ehm, if it is not going to be a straight line, what else?
\end{quote}

In this exchange, the students begin by using neighborhood analysis (EB~ID) and making sense of the behavior of the graph at the boundaries (PMA~5). They reason about the rate of change by discussing the physical nature of the quantities. They then connect the points by choosing a function based on the trend of the quantities and the physical context (function choosing, EB~IIB). The students agree that a linear relationship is likely appropriate here. Like the math students discussed above, a linear relationship between quantities is a default assumption for these physics students. However, the physics students are continually relating their representation to the physical quantities and context. Later in their discussion, Evan continues to argue for why the function should be linear, based on the physical properties of the rod:

\begin{quote}
    Evan: Cause, yeah, if it wouldn't be a straight line, then it should be something else and I cannot imagine what it would, would be then. Because the rod in its whole is heat conductive in the same way everywhere. So, if this side is held perfectly at zero and that side is perfectly held at 50 (adds ``0'' to the starting point of the three graphs and ``50'' to the end point of the three graphs) and everything in between is heat conductive in the same way, it doesn't seem to me that it would be something else than a straight line\ldots
\end{quote}

We note that the students here exhibit journeyman expertise---that of one between a novice and a true expert \cite{Bing2012}---in rejecting non-linear models based on physical sensemaking with quantity.

Analysis of these results from physics education research for both introductory and upper-division through the lens of the CoRP framework reveal the importance of mathematization and and understanding of quantity for productive physics covariational reasoning. At the introductory level, students seem to display reasoning that is evidence of understanding of physics quantities as an emerging resource, reasoning about some---but not all---quantities across multiple contexts. We suggest that encouraging sensemaking about quantities could support development of physics covariational reasoning at this stage. The analysis of the upper-division students' work using the CoRP framework illuminates the importance of understanding of quantity as a foundation of physics covariational reasoning in more sophisticated contexts: the upper-division students' understanding of the physics context and relevant quantities allows the students to productively engage in the expert behavior of Function Choosing. 

In summary, in this section we have demonstrated how the CoRP framework facilitates recognizing several emerging resources that are important for helping calculus and  physics students develop their mathematical reasoning in physics: 

\begin{enumerate}
    \item Introductory students recognize that covarying quantities are related by function; an understanding of the quantities is helpful for reasoning about the functions involved. Students' familiarity with linear functions could help in developing reasoning skills about other functions, which are more challenging at this level. 
    \item Upper-division students are familiar with more physics quantities, and have a deeper understanding of some. Relying on student understanding of quantity might help further their skills at this level to choose between different symbolic models---both linear and non-linear.
\end{enumerate}

\section{Conclusion}  

Physics covariational reasoning plays a central role in expert-like quantitative modeling and is a key aspect of what it means to ``think like a physicist.'' Because of its focus on quantitative modeling, introductory-level physics provides a unique opportunity for the development of physics covariational reasoning for a large population of students. Therefore, reliable covariational reasoning with physical quantities is a desirable student learning outcome of introductory physics courses. This is especially true because quantitative reasoning developed by instruction in physics is transferable to ``real-world'' contexts, and provides a foundation for scientific literacy more generally. However, it can be challenging for instructors to recognize the ways in which their own reasoning patterns differ from the focus of the prerequisite mathematics their students have taken. The differences in mathematics and physics covariational reasoning was one significant motivation for the development of the CoRP framework.

The use of covariational reasoning in physics contexts has not before been operationalized. The CoRP framework provides one operationalization. We expect that it will help instructors recognize the role (and importance) of covariational reasoning in quantitative modeling in physics, and better address its development as an important student learning outcome. In this paper, we have described three ways that the CoRP framework can be used to achieve this goal.

First, the CoRP framework can make clear the ways in which physics covariational reasoning is distinct from the covariational reasoning described by mathematics education researchers and taught in mathematics courses. Frameworks developed by the mathematics education research community focus largely on direct consideration of how changes in one variable results in changes to another. While consideration of changing quantities plays an important role in physics covariational reasoning, the interconnected structure and facets of the CoRP framework attend to research that demonstrates the inherently blended nature of mathematical and physical reasoning \cite{Eichenlaub2019, VanDenEynde2019, Serbin2022TheProblems, strawberryfields}. The foundation of physics covariational reasoning is the Proceptual Understanding of the underlying mathematics and physics mathematization. The Physics Mental Actions described in the CoRP framework are similar to the mental actions described by Carlson et al. \cite{carlson2002}, but also involve quantities, rather than variables which may be free of physical context. We argue that consideration of how quantities change with respect to each other cannot happen effectively without understanding of the quantities themselves. The expert behaviors in the CoRP framework also rely on physics content knowledge, and often guide use of the Physics Mental Actions. 

Students may come into introductory physics courses with experience with covariational reasoning in math contexts, but as shown in section \ref{sec:CoRPrelevance}, this does not guarantee facility with physics covariational reasoning. The characterization of physics covariational reasoning provided by the CoRP framework provides guidance for leveraging the experience that students have from mathematics courses. Familiarity with how covariational reasoning is used in physics, as described in the framework, can help instructors meet their students where they are, leading to more productive reasoning and quantitative modeling in physics contexts.

Second, the CoRP framework can be used to analyze student reasoning in a variety of contexts and in a number of ways; here, we discuss two such ways. First, analysis of introductory student covariational reasoning can help instructors track changes in students' covariational reasoning that occur with instruction in physics, and can inform assessment, as it explicates facets of covariational reasoning. This allows covariational reasoning to be a truly \emph{assessable} learning objective of introductory physics courses. Also, analysis of student reasoning can aid education researchers and curriculum developers. As demonstrated in Section \ref{sec:CoRPrelevance}, analysis of students' reasoning through the lens of the CoRP framework allows for the identification of expert-like and proto-expert-like covariational reasoning, as well as aspects of covariational reasoning in students' zones of proximal development. 

Finally, the CoRP framework can guide the development of instructional interventions. The framework identifies proceptual understanding of the foundational mathematics and physics mathematization as a basis of physics covariational reasoning. Covariational reasoning in physics is not simply doing math with physics quantities---it requires deep understanding and facility with both the mathematics and the quantities themselves. By identifying the foundations of physics covariational reasoning, the CoRP framework provides a way to determine ``essential skills'' \cite{mikula2017} that can be targeted with interventions or instruction. For example, while students may come into physics courses able to produce or interpret graphs in purely mathematical contexts, they may lack facility with the physics knowledge embedded in a graph of physics quantities. Being able to interpret the meaning of a graphical feature such as a slope or an area under a curve, or identify a quantity associated with a graphical feature, may aid physics learners in understanding graphical representations. 

Student difficulties with mathematics in a physics context have long been viewed as a problem of mathematical under-preparedness. While lack of adequate practice with algebraic manipulations characterizes some students' difficulties, physics has its own work to do in helping its students learn to reason mathematically in a physics context. The work described in this paper is situated in ongoing efforts to help build physics quantitative literacy for all physics students. We have developed the CoRP framework to help support the research and instructional communities in physics gain new knowledge and develop instructional interventions. We suggest the CoRP framework can help scaffold physics students' understanding of mathematical models through their development of covariational reasoning.

\acknowledgements{This work is supported by the National Science Foundation under grants No. DUE-1832836, DUE-1832880, DUE-1833050, DGE-1762114. The work described in this paper was performed while the first author held an NRC Research Associateship award at Air Force Research Laboratory.
}

\clearpage

\bibliography{covar.bib}
\end{document}